\documentclass[10pt,prb,twocolumn,amsmath]{revtex4}
\newcommand{\eq}[1]{Eq.~{(\ref{#1})}}
\newcommand{\fig}[1]{Fig.~{\ref{#1}}}
\newcommand{\bea}{\begin{eqnarray}}
\newcommand{\beann}{\begin{eqnarray*}}
\newcommand{\eea}{\end{eqnarray}}
\newcommand{\eeann}{\end{eqnarray*}}

\usepackage{graphicx}
\usepackage{latexsym}
\usepackage{amsbsy}
\newcommand{\bs}{\boldsymbol}
\renewcommand{\vec}[1]{\bs{#1}}

\begin{document}
\title{Geodesics in the Generalized Schwarzschild Solution}
\author{Matthew R. Francis} \email{mfrancis@physics.rutgers.edu}
\author{Arthur Kosowsky} \email{kosowsky@physics.rutgers.edu}
\affiliation{Dept.\ of Physics and Astronomy, Rutgers University\\136
Frelinghuysen Road, Piscataway, NJ 08854} \date{\today}

\begin{abstract}
Since Schwarzshild discovered the point-mass solution to Einstein's
equations that bears his name, many equivalent forms of the metric
have been catalogued.  Using an elementary coordinate transformation,
we derive the most general form for the stationary,
spherically-symmetric vacuum metric, which contains one free function.
Different choices for the function correspond to common expressions
for the line element.  From the general metric, we obtain particle and
photon trajectories, and use them to specify several time coordinates
adapted to physical situations.  The most general form of the metric
is only slightly more complicated than the Schwarzschild form, which
argues effectively for teaching the general line element in place of
the diagonal metric.
\end{abstract}
\maketitle

\section{Introduction}

For many decades, general relativity was considered a highly abstract subject,
the purview of sophisticated mathematical physics far removed from the
undergraduate curriculum. This situation has changed in recent years
as general relativity has matured into a true experimental science:
gravitational lensing now routinely probes dark matter in distant
galaxies\cite{lensing}, gravitational wave detection thresholds are steadily
marching into the range of realistic sources\cite{ligoPT,periodic}, and the Global
Positioning System\cite{gps} must compensate for gravitational time delays. In
parallel to this evolving status of general relativity, a number of
excellent textbooks at the undergraduate level have been published in
recent years\cite{ellis,schutz,foster,hartle}, demonstrating that
teaching general relativity to physics majors is not only possible but
exciting and relevant.

One of the elementary results of general relativity is the spacetime
arising from a point mass $M$, which is spherically symmetric
and \emph{stationary}, or independent of the time coordinate.
Close to the mass, the spacetime represents a black hole with an event
horizon, while at a large distance it reproduces the results of
Newtonian gravity. This spacetime is most commonly represented
by the Schwarzschild metric
\begin{subequations}
\bea
ds^2 = g_{\mu\nu} dx^\mu dx^\nu = \chi(r) dt^2 - \chi(r)^{-1} dr^2 
- r^2 d\Omega^2,
\eea
\bea
\chi(r) \equiv 1 - \frac{2 M}{r},
\label{chidef}
\eea
\label{schwarzschild}
\end{subequations}
where we have used units $c = G = 1$ and abbreviated $d\Omega^2 \equiv 
d\theta^2 + \sin^2 \theta d\phi^2$. Historically, this was
the first widely-used form of the metric\footnote{Antoci and Liebscher \cite{antoci} have recently noted that
Schwarzschild's original derivation\cite{schwarz_trans} used 
different coordinates, and the familiar form bearing Schwarzschild's name 
is actually due to Hilbert.}, and virtually all texts begin (and
often end) the discussion of spherically symmetric metrics
with \eq{schwarzschild}. Formally, this form of the metric is
in one sense the simplest possible, since it is the only diagonal form
of the metric, and this property
results in simplified calculations (e.g.,
the smallest possible number of non-zero connection coefficients).

However, metrics in general relativity are not physically measurable
quantities, and numerous metrics employing different sets of coordinates
can be used to describe the same
physical spacetime. The usual form for the Schwarzschild metric
possesses some peculiar properties which make it far from ideal
pedagogically. Most notably, it contains an unphysical coordinate
singularity at the event horizon $r=2 M$, the interpretation
of which is confusing enough that its true nature was debated for
many years.  A second more subtle
point is that the metric is the same under time-reversal: a particle
traveling forward in time is equivalent to one traveling backward in
time, but physically particles can cross the event horizon in only one
direction.

Here we perform a general coordinate transformation of
Eq.~\ref{schwarzschild} to obtain the most general form of the 
stationary, spherically-symmetric metric:
\bea 
ds^2 = \chi(r) du^2 &+& 2 B(r) du dr \nonumber \\ 
&-& \chi^{-1}(1-B(r)^2) dr^2 - r^2 d\Omega^2 .
\label{GLE}
\eea
This form contains one free function of radius, $B(r)$, which explicitly
displays the coordinate freedom in the metric. Through various choices
for $B(r)$, all of the familiar (and not-so-familiar) forms of
the metric can be obtained, including the Schwarzschild form and
the Eddington-Finkelstein metric often used to investigate the physical
nature of the event horizon. We advocate the use of \eq{GLE} as a much
clearer and pedagogically versatile form of the metric for a number
of reasons:
\begin{enumerate}
\item \eq{GLE} emphasizes the fact that the same physical spacetime is
described by many different metrics;
\item It unifies the disparate expressions
for the metric in the general relativity literature into a single
simple form;
\item It naturally suggests variables describing motion
of particles which are closely tied to physically measurable quantities,
and investigating the reduction to different metric forms clarifies the
spurious nature of the singularity in \eq{schwarzschild}; and
\item Mathematical manipulations involving \eq{GLE} are only
slightly more complex than those using the Schwarzschild form.
\end{enumerate}
While \eq{schwarzschild} may be the simplest form of the
metric, it does not give the simplest or most physically 
transparent forms for equations of particle
motions in the spacetime.

The rest of this paper derives the general form of the metric, demonstrates
its reduction to various specific forms, and discusses the 
corresponding expressions for the geodesic paths followed by 
light and particles moving in the radial direction.
In the next Section, \eq{GLE} is derived from the familiar
\eq{schwarzschild}  via a straightforward coordinate transformation.
Section~\ref{radial_geod} establishes expressions for radial geodesics in the
general case, and Sec.~\ref{coords} examines how the geodesic expressions
are simplified for specific choices of the arbitrary function $B(r)$,
including all of the most commonly used forms of the metric.
Three brief appendices cover relevant but somewhat more technical material:
Appendix~\ref{eesoln} gives the connection and Ricci curvature components
associated with \eq{GLE} for convenience in doing calculations; 
Appendix~\ref{standard_tests} discusses why a change in time
coordinate does not affect the gravitational redshift; and
Appendix~\ref{vierbeins} expresses the general line element in
vierbein form. The derivations and arguments in the body of the
paper require no mathematics beyond elementary calculus, and as
such are suitable for inclusion in any undergraduate general relativity
class.

\section{The General Line Element}
\label{tgle}

In the Schwarzschild form of the metric, \eq{schwarzschild},
the time direction is orthogonal to all other directions (that is,
$g_{0k} =0$ when $k\neq 0$) so the metric is called \emph{static};
\eq{schwarzschild} is the unique static spherically-symmetric vacuum
line-element.  The radial coordinate $r$ in the Schwarzschild
geometry has a clear physical interpretation: it
is the proper distance measured by a massive particle falling
into a black hole\cite{gautreau}. The time coordinate $t$ is
more problematic. The proper time for a particle
(either massive or massless) falling through the event horizon is
finite, but the coordinate time $t$ diverges---it takes an infinite amount
of time in the Schwarzschild system of coordinates to fall into a
black hole\cite{foster}. Thus the singularity of \eq{schwarzschild}
at the point $r=2M$ is not physical, and can be transformed away
with a different choice of coordinates.
We are interested in forms of the metric that are
stationary but are not necessarily diagonal.

A transformation of the time coordinate of the type
\bea
u = \alpha t + \beta(r),
\label{time_transform}
\eea 
for some function $\beta$ and constant $\alpha$, ensures that the
metric remains stationary, while keeping the spherical symmetry
manifest.  In other words, if $\alpha$ and $\beta$ remain
unspecified, we obtain the most general stationary vacuum line
element in spherical coordinates.  Without loss of generality we
take $\alpha = 1$; a simple rescaling of the time variable
is equivalent to choosing a different time unit.
Since $u$ is a function of $t$ and $r$,
its differential is 
\bea du &=& \frac{\partial u}{\partial t} dt +
\frac{\partial u}{\partial r} dr 
= dt + \frac{\partial \beta}{\partial r} dr \nonumber \\ 
du^2 &=& dt^2 + 2 \beta'(r) dt dr + \beta'(r)^2 dr^2 .
\label{du}
\eea

Rather than going through the complicated algebra of solving for $dt$
and substituting into \eq{schwarzschild}, note that a
general stationary line element must have the form 
\beann 
ds^2 = A(r) du^2
+ 2 B(r) du dr - C(r) dr^2 - r^2 d\Omega^2 ; 
\eeann 
setting this equal
to \eq{schwarzschild} and using the relations in \eq{du},
\begin{subequations}
\bea
A(r) &=& \chi(r), \\
B(r) &=& - \chi(r) \beta'(r), \label{betaelim}\\
C(r) &=& \chi(r)\beta'(r)^2 + 2 B(r) \beta'(r) + \chi(r)^{-1}. \label{C}
\eea
\end{subequations}
We can use \eq{betaelim} to eliminate $\beta' = - B(r)/\chi(r)$, 
which leads to a condition on $B$ and $C$ via \eq{C}:
\bea
C(r) \chi(r) + B(r)^2 = 1 ,
\label{gauge}
\eea 
and \eq{GLE} follows.  As Appendix~\ref{eesoln}
shows explicitly, the Einstein field equations do not specify $B(r)$, so we
will call \eq{GLE} the general line element for the
spherically-symmetric vacuum equations.  We are free to choose $B(r)$: 
for example, the Schwarzschild form of the metric is
obtained from the general line element by letting $B(r) = 0$.  We will
discuss other coordinate choices below.  
Appendix~\ref{standard_tests} examines how a change in time coordinate,
specified by $B(r)$ and $\chi(r)$,
leaves some of the standard tests of general relativity
unaffected.

Of course, it is also possible to change the spatial coordinates
$(r,\theta,\phi)$, but it is nice to leave them as written for two
reasons: spherical symmetry is explicit, and as mentioned in the
Introduction, the radial coordinate $r$ is a measure of the proper
distance measured by a massive particle falling radially into the
black hole.

\section{Radial Geodesics}
\label{radial_geod}

Since free particles (including photons) follow geodesics in general
relativity, the Lagrangian for free motion can be written simply as
the squared norm of the velocity $\dot{\vec{x}} = d\vec{x}/d\lambda$
with respect to some parameter $\lambda$\cite{foster}: 
\bea 2
\mathcal{L}(\vec{x},\dot{\vec{x}},\lambda) &=& g_{\mu\nu}
(\vec{x})\dot{x}^\mu \dot{x}^\nu = \chi \dot{u}^2 
+ 2 B \dot{u} \dot{r} \nonumber \\ 
&-& \chi^{-1} (1- B^2) \dot{r}^2 - r^2 (
\dot{\theta}^2 + \sin^2 \theta \dot{\phi}^2 ) .  
\eea 
The tests of
general relativity that depend on the angular variables (light
deflection, for example) do not explicitly involve the time
coordinate, so we will look strictly at radial geodesics by letting
$\theta = \pi/2$ and $\dot{\phi} = 0$.

The magnitude of a geodesic vector (and hence the Lagrangian itself)
is a constant of motion, so we have 
\bea g_{\mu\nu} \dot{x}^\mu
\dot{x}^\nu = \chi \dot{u}^2 + 2 B \dot{u} \dot{r} - \chi^{-1}(1-B^2)
\dot{r}^2 \equiv \kappa
\label{kappa}
\eea where $\kappa = 1$ for massive particles and $\kappa = 0$ for
photons.  The time coordinate $u$ is cyclic, so we know that 
\bea
\frac{\partial\mathcal{L}}{\partial \dot{u}} = \chi \dot{u} + B \dot{r} \equiv
\epsilon
\label{energy}
\eea 
is constant, with $\epsilon$ related to the energy of the
particle (with respect to an observer at infinity) as follows: 
\beann
\epsilon = \left\{ 
\begin{array}{lll} E/m &\quad& \text{massive
particles} \\ E/\sqrt{\hbar} &\quad& \text{photons} 
\end{array}
\right.  
\eeann 
Solving \eq{energy} for $\dot{u}$ and substituting it
into \eq{kappa} allows us to solve for $\dot{r}$: 
\beann \dot{r}^2 = \epsilon^2 - \kappa\chi .  
\eeann 
Since this is independent of the
choice of $B$, the radial geodesic equation is as well: 
\bea \ddot{r}
+ \kappa M/r^2 = 0, 
\eea 
which is the Newtonian expression (upon
restoration of $G$)---although $r$ is not the flat spacetime radial
coordinate, and when $\kappa = 1$, $\lambda$ is not the 
coordinate time but the proper time shown by clocks moving
with the massive particle along the geodesic\cite{foster}.

Using our expression for $\dot{r}$ in \eq{energy} 
allows us to write the velocities with respect to the proper time:
\begin{subequations}
\bea
\dot{r} &=& w \sqrt{ \epsilon^2 - \kappa \chi} \label{rdot}\\
\dot{u} &=& \chi^{-1} 
\left\{ \epsilon - w B \sqrt{\epsilon^2 - \kappa \chi } \right\} \label{udot}
\eea
\label{rudot}
\end{subequations}
where $w = + 1$ for outgoing geodesics, while $w = -1$ for incoming.
From these, we can find the coordinate time as a function of radius,
\bea 
u(r) = \int \frac{\dot{u}}{\dot{r}} dr = \int \frac{dr}{\chi}
\left\{ \frac{w}{\sqrt{1 - \kappa \chi/\epsilon^2}} - B \right\}, 
\eea
which tells us whether a particle crosses the event horizon in finite
coordinate time, for a particular choice of $B(r)$.  In the
Schwarzschild case $B = 0$, for example, the time required diverges
logarithmically for both values of $w$\cite{foster}.  Once $B$ is
chosen, two parameters dictate the type of geodesic: $\kappa=0$ or 1, which
specifies lightlike (photons) versus timelike (massive particles)
geodesics, and $\epsilon$, which
determines the energy of the particle.  In the case of null geodesics
($\kappa = 0$), the photon energy is irrelevant: 
\bea u(r) = \int
\frac{dr}{\chi} \left( w - B \right).
\label{urnull}
\eea

\section{Coordinate Systems from Geodesics}
\label{coords}

In the following subsections, we will examine choices of $B(r)$
that simplify the geodesic expressions.  Some of these examples are
familiar, others less so. Our central conceptual point is that
the general form of the metric, \eq{GLE}, allows us to choose
a specific system of coordinates which leads to simple and
physically intuitive forms for the \emph{geodesics}. Since the
geodesics, unlike the metric, represent physically observable
quantities, picking a coordinate system based on the properties
of the geodesics instead of the metric makes sense, as long as
the coordinate system does not lead to unduly complex computations.

\subsection{Two Simple Systems}
\label{two_systems}

Equation~(\ref{urnull}) gives the coordinate time as a function of
radius for a null geodesic.  Future-pointing paths of light in special
relativity are straight lines defined by $u(r) = - r + const.$; to
replicate this, we let $B = \chi - 1 = -2 M/r$, which means that
\beann
u_{EF}(r) = \left\{ \begin{array}{lll} -r + u_0 &\quad& \text{incoming} \\ 
r + 4 M \ln | r - 2 M | + u_0 &\quad& \text{outgoing} \end{array} \right.
\eeann
where $u_0$ is a constant of integration.  This is one version of the
familiar Eddington-Finkelstein coordinates; ingoing geodesics are
straight lines, while outgoing must ``climb out'' of the gravity well
formed by the black hole.  Nothing starting within the event horizon
can get out, since the ``outward'' geodesics starting from $r < 2 M$
actually fall inward as well.

The line element for this choice of coordinates is 
\beann ds^2_{EF}
&=& du^2 - dr^2 - r^2 d\Omega^2 - \frac{2 M}{r} \left( du - dr \right)^2 \\ 
&=& \left(\eta_{\mu\nu} - 2 V l_\mu l_\nu \right) dx^\mu dx^\nu, 
\eeann 
where in the second line we have written the metric as the sum of the flat
metric (in spherical coordinates) and a contribution from a null
vector 
\beann \left[l^\mu\right] = (1,- 1,0,0).  
\eeann 
Metrics of
this type are often called \emph{Kerr-Schild} metrics, and in their
more general form are used to find the rotating black hole solution (known as the Kerr solution; see Ref.~[\onlinecite{misner}] for more details).

A related class of coordinates can be obtained by setting the proper
time (for a massive particle) or affine parameter (for a photon)
equal to the coordinate time for an inbound geodesic.  This means we
must specify \emph{in advance} the energy $\epsilon'$ of the particle,
and whether it is massless or massive by setting the parameter
$\kappa'$:
\beann
\dot{u} = 1 = \frac{\epsilon'}{\chi} 
\left\{ 1 + B \sqrt{1-\kappa' \chi/\epsilon'{}^2 } \right\}.
\eeann
Selecting the retarded solution, we find
\bea
B = \frac{1}{\sqrt{1-p\chi}} \left\{ \chi \sqrt{\frac{p}{\kappa'}} - 1 \right\}
\label{geod_time}
\eea
where
\beann
p \equiv \kappa'/\epsilon'{}^2 
\eeann
is the metric parameter that determines the time coordinate associated
with a particular energy.  When $\kappa' = p = 0$, $B = \chi/\epsilon'
-1$; if $\epsilon' \to 1$ (so that $|\dot{r}| = 1$), we get back the
Eddington-Finkelstein time coordinate.  Thus, despite the appearance
of both $p$ and $\kappa'$ in \eq{geod_time}, they are \emph{not}
independent parameters.  Since we have already discussed the
Eddington-Finkelstein case, we will consider only $\kappa' = 1$ and 
$p \neq 0$ for the remainder of the section.

The line element for this geometry with $\kappa' =1$ is
\bea
ds^2_{GT} = \chi du^2 &+& 
2 \frac{\chi \sqrt{p} -1}{\sqrt{1-p\chi}} du dr \nonumber \\ 
&-& \frac{2 \sqrt{p} - p (1 + \chi)}{1 - p \chi} dr^2  - r^2 d\Omega^2 
\label{GTelem}
\eea
where the subscript ``GT'' indicates geodesic time.  For an arbitrary
geodesic, the trajectory is described by
\beann
u_p(r) &=& \int \frac{dr}{\chi} 
\left\{ \frac{w}{\sqrt{1-\kappa\chi/\epsilon^2}} 
- \frac{\chi \sqrt{p} - 1}{\sqrt{1 - p \chi}} \right\}
\eeann
Plots for different values of $p$ and $\epsilon$ are shown in
\fig{GT-LMP}a.  When $\epsilon = \epsilon'$ (and $\kappa = 1$), the
time for an infalling geodesic takes the simple form
\beann
u_p(r) &=& \int dr \sqrt{ \frac{p}{1-p\chi}}
\eeann
which reduces to the Eddington-Finkelstein case in the limit $p\to 0$.

The relative complexity of the line element in \eq{GTelem} may be why
it is not more widely known---except in the specific case $p = 1$,
which is the Painlev\'{e}-Gullstrand metric (see
Ref.~\onlinecite{martel}), independently discovered by Lasenby, Doran,
and Gull\cite{GTG}:
\bea
ds^2_{PG} = \chi du^2 + 2 \sqrt{\frac{2 M}{r}} du dr - dr^2 - r^2 d\Omega^2 .
\label{PG}
\eea
We will see this metric again as a special case in the following section.

\subsection{Relationship Between $u$ and Proper Time}
\label{LMP_coords}

In the previous section, we have chosen coordinate
times $u$ which lead to physically intuitive forms for radial geodesics,
specified by $u(r)$. In one case we explicitly equate the coordinate
time with the proper time for a massive particle moving along a radial
geodesic, \eq{GTelem}. More generally, we can derive the proper time $u$
for any choice of coordinate time $u$ (or equivalently, the function
$B(r)$). 
Starting from \eq{kappa}, we multiply both sides by $d\lambda/dr$ 
and rearrange somewhat:
\bea
\kappa' \frac{d\lambda}{dr} = \left( \chi \frac{du}{dr} 
+ 2 B \right) \frac{du}{d\lambda} 
- \frac{1}{\chi}(1 - B^2) \frac{dr}{d\lambda}.
\label{step1}
\eea
Applying \eq{udot} to the first term of the right-hand side eliminates 
$du/d\lambda$
\beann
\left( \chi \frac{du}{dr} + 2 B \right) \frac{\dot{u}}{\epsilon'} 
= \frac{du}{dr} - \frac{du}{dr} w B \sqrt{1 - p \chi} \\ 
+ \frac{2 B}{\chi} (1 - w B \sqrt{1 - p\chi} ) \\
= \frac{du}{dr} + B \left( 2 \chi^{-1} - 1 \right) 
\left( 1 - w B \sqrt{1 - p \chi} \right),
%\label{step2}
\eeann
where $p \equiv \kappa'/\epsilon'{}^2$ as before, and we have used the
identity $du/dr = \dot{u}/\dot{r}$.  Substituting \eq{rdot} and
integrating both sides of \eq{step1} over $r$ yields
\bea
\epsilon' p \lambda = u + \int \frac{dr}{\chi} 
\left( B - w \sqrt{1 - p \chi}\right),
\label{proper_transform}
\eea
which is a coordinate transform as in \eq{time_transform}. 
For massive particles with $p\neq 0$, 
\eq{proper_transform} gives the proper time
$\lambda$ in terms of the coordinate time $u$; when
$p=0$, no proper time can be defined but
the integral is the same as the one in \eq{urnull}. 

\begin{figure*}[pt]
	\includegraphics[scale=0.6]{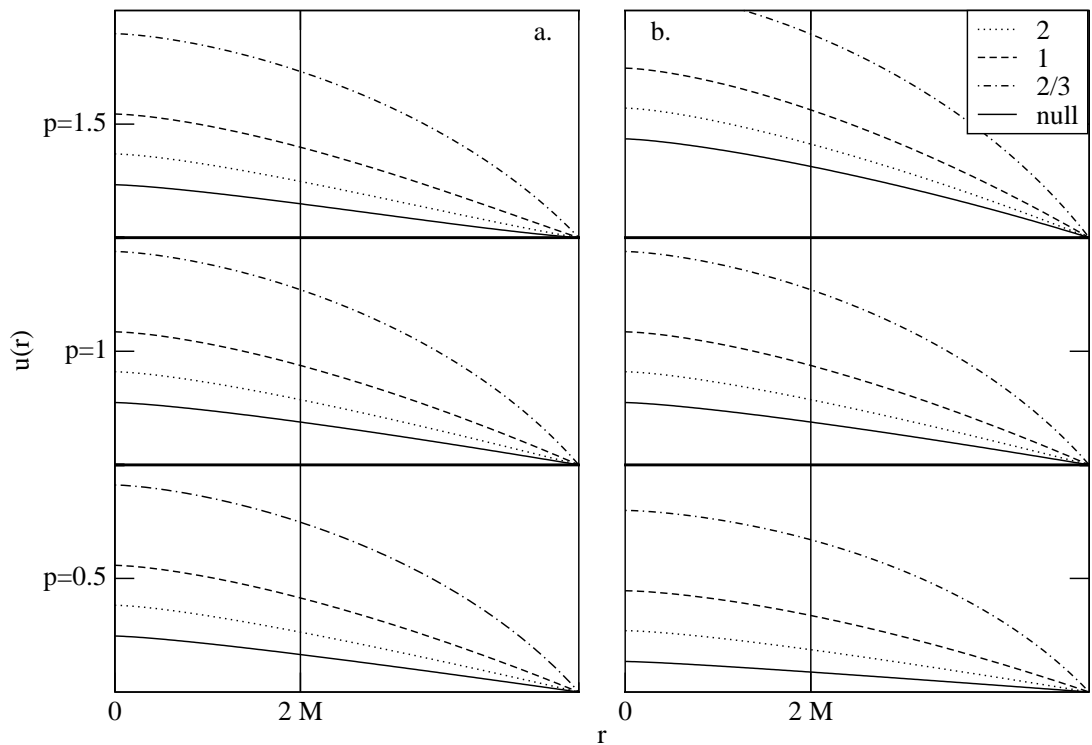} 
\caption
{a. Plot of the geodesic time parameter $u_{GT}(r)$ for various
values of $p$.  The legend indicates different values for the
energy parameter $\epsilon$, as well as the null geodesic.  
b. Plot of the Lake-Martel-Poisson time parameter $u_{LMP}(r)$
under the same conditions as the geodesic time.  Note that the
times are identical ($u_{GT}=u_{LMP}$) for $p=1$.}
\label{GT-LMP}
\end{figure*}

For special choices of $B$, the integral in \eq{proper_transform}
takes a simpler form.  The obvious example is
\bea
B = - \sqrt{1 - p \chi},
\label{lmp}
\eea
which gives rise to the Lake-Martel-Poisson\cite{lake,martel} metric
\beann
ds^2 = \chi du^2 - 2 \sqrt{1 - p \chi} du dr - p dr^2 - r^2 d\Omega^2 .
\eeann
Different values of $p$ yield several well-studied solutions: when
$p=0$, we obtain the more familiar form of the retarded
Eddington-Finkelstein metric\cite{foster} (in which infalling null
geodesics are lines of constant $u$), while $p=1$ yields the
Painlev\'{e}-Gullstrand metric of \eq{PG}.  We may also treat $p$ in
general as a positive parameter (which can be associated with the
energy of a particle at infinity\cite{martel})in which case $r$ can
range from zero to infinity for $0 \leq p \leq 1$, and
\beann
r \leq 2 M p/(p-1)\quad  \text{for} \quad p > 1 .
\eeann
The latter case is proportional to the Gautreau-Hoffmann coordinate
system\cite{gautreau}, with the initial radial value defined by 
$R_i = 2 M p/(p-1)$ in our notation.  Note that \eq{lmp} is equivalent to
\eq{geod_time} when $p = \epsilon' = 1$; in this limit, by construction,
the proper time is equal to the coordinate time.  This case is equivalent
to setting $\dot{u} = \epsilon' p$ and repeating the analysis of
Section~\ref{two_systems}.  Plots for different values of $p$ and
$\epsilon$ are given in \fig{GT-LMP}b.

\section{Discussion}

The general line element provides a clear and 
straightforward way to pick specific time
coordinates for the spherically symmetric, stationary spacetime.  
\eq{GLE} is a simpler starting point than the diagonal
Schwarzschild metric for determining the two forms of the
Eddington-Finkelstein coordinate system, and provides both the
geodesic time parameter and the Lake-Martel-Poisson coordinates directly,
as opposed to performing a rather complicated coordinate
transformation.  In fact, the geodesic time line element is new to us,
and obtaining it directly from the Schwarzschild solution involves a
difficult integral.

We have emphasized stationary forms of the metric; however, it is
possible to ``maximally continue'' any of these coordinate systems to
make them diagonal.  This is done by using
\beann
u \equiv t + \beta(r)\quad \text{and} \quad v \equiv t - \beta(r)
\eeann
in place of $t$ and $r$, and rewriting the line element in the new
coordinates.  The familiar example of this is the Kruskal-Szekeres
extension, which uses the Eddington-Finkelstein coordinates as its
starting point.  However, all maximally-extended solutions include
a non-physical region of spacetime (the
``white hole'' problem), and this poses as serious an interpretational 
difficulty as the coordinate singularity problem at the event horizon.

From a pedagogical point of view, we emphasize that \eq{GLE} contains
virtually all other discussed coordinate systems as special cases.
The general metric is only slightly more
complicated to use in calculations than the diagonal Schwarzschild
case; for convenience we list the connection and Ricci tensor
components explicitly in Appendix~\ref{eesoln}. For reasons
of pedagogical and conceptual clarity, We advocate presenting
\eq{GLE} as the line element describing the spherically symmetric
and stationary vacuum spacetime when teaching general
relativity, especially at the undergraduate level;
the familiar Schwarzschild form of the metric can be easily
obtained as a special case. 
Appendix~\ref{vierbeins} shows how to derive \eq{GLE} in a
straightforward way which is no more complicated than
usual derivations of the Schwarzschild metric.

\begin{acknowledgments}

This work is partly supported by NASA through grant NAG5-10110.  AK is
a Cottrell Scholar of the Research Corporation.  We thank Ronald Byers
for his careful reading of the manuscript, and two anonymous referees
for their comments.

\end{acknowledgments}

\begin{appendix}

\section{Solution of Einstein's Equations}
\label{eesoln}

The Christoffel connection coefficients are found using the standard formula
\beann
\Gamma^\lambda_{\mu\nu} = \frac{1}{2} g^{\lambda\kappa} 
\left( \partial_\mu g_{\kappa\nu} + \partial_\nu g_{\kappa\mu} 
- \partial_\kappa g_{\mu\nu} \right).
\eeann
If we take \eq{GLE} to be an ansatz for solving Einstein's equations, 
with unknown functions $\chi(r)$ and $B(r)$, the nonzero connection 
coefficients are
\beann
\Gamma^0_{00} &=& - \Gamma^1_{01} = B \chi'/2 \\
\Gamma^0_{01} &=& - \Gamma^1_{11} = C \chi'/2 \\
\Gamma^0_{11} &=& C B' - B C'/2 \\
\Gamma^0_{22} &=& \Gamma^0_{33} \csc^2\theta = B r \\
\Gamma^1_{00} &=& \chi \chi'/2 \\
\Gamma^1_{22} &=& \Gamma^1_{33} \csc^2\theta = -\chi r \\
\Gamma^2_{12} &=& \Gamma^3_{13} = r^{-1} \\
\Gamma^2_{33} &=& -\sin\theta\cos\theta \\
\Gamma^3_{23} &=& \cot \theta
\eeann
where we have used the shorthand (see \eq{gauge})
\beann
C = \chi^{-1}(1-B^2)
\eeann
and $\chi' = d\chi/dr$.  These expressions are not complicated, 
but there are more of them than in the diagonal case.

The Ricci tensor components,
\beann
R_{\mu\nu} = \partial_\nu \Gamma^\sigma_{\mu\sigma} 
- \partial_\sigma \Gamma^\sigma_{\mu\nu} 
+ \Gamma^\rho_{\mu\sigma}\Gamma^\sigma_{\rho\nu} 
- \Gamma^{\rho}_{\mu\nu}\Gamma^{\sigma}_{\rho\sigma} ,
\eeann
turn out to be quite simple:
\beann
R_{00} &=& \chi \Phi/2 r \quad
R_{01} = B \Phi/2 r \quad
R_{11} = - C \Phi/2 r \\
R_{22} &=& R_{33} \csc^2\theta = 1-\chi - \chi' r
\eeann
where
\beann
\Phi \equiv \chi'' r + 2 \chi' .
\eeann
Thus, Einstein's vacuum equations $R_{\mu\nu} = 0$ do not involve $B(r)$, 
and are easily solved to give \eq{chidef}, with $M$ as an integration
constant which is identified with the total mass by ensuring
the Newtonian limit.

\section{New Time Variables and the ``Standard Tests''}
\label{standard_tests}

Most derivations of the measureable effects due to a planet, star, or
black hole start from \eq{schwarzschild}; nevertheless, changing time
variables does not change the predictions of general relativity.  We
will not bother to rederive all the familiar expressions, but instead
will emphasize their independence of particular coordinate
assumptions.  In fact, most tests of GR (bending of light, perihelion
advance of an object in orbit, etc.) depend on the radial velocity
with respect to proper time, which is a generalization of \eq{rdot}.
Since this equation does not involve $u$ in any way, we don't need to
discuss these tests here.  The exceptions are time delay and
gravitational redshift; demonstrating that the former is independent
of time coordinate choice requires coverage of non-radial motion, so
we will not discuss it here.

For gravitational redshift, let $[\xi^\mu ] \equiv (1,0,0,0)$ be a 
timelike (Killing) vector in whatever coordinate system we desire.  
The ``direction of time'' at a given value of $r$ is then indicated by
\bea
\hat{\xi}^\mu (r) = \xi^\mu /\sqrt{\xi^\mu \xi_\mu} = \xi^\mu /\sqrt{\chi(r)}.
\label{unitvec}
\eea
If we locate two stationary observers $O_1$ and $O_2$ at $R_1$ and
$R_2$, respectively, and send a pulse of light between them, we obtain
a shift in frequency due to the presence of $\chi$.  The path of the
light is indicated by the null momentum vector $\vec{k}$, whose time
component gives the frequency:
\beann
\omega_1 = \left. k_\mu \hat{\xi}^\mu \right|_{r=R_1}, \ \omega_2 = 
\left. k_\mu \hat{\xi}^\mu \right|_{r=R_2} .
\eeann
The projection of the timelike vector $\vec{\xi}$ along the geodesic
is constant (although proof of this conjecture requires a discussion
of Killing vectors\cite{hartle,wald}, which is beyond the scope of
this paper).  Using \eq{unitvec}, we obtain
\beann
\omega_1 = k_\mu \xi^\mu /\sqrt{\chi(r_1)}
\eeann
and a similar expression for $\omega_2$.  Taking the ratio eliminates 
the dependence on $\vec{k}$, leaving us with
\bea
\frac{\omega_1}{\omega_2} = \sqrt{\frac{\chi(r_2)}{\chi(r_1)}}.
\eea
The redshift formula is directly obtained from this ratio; since it
doesn't depend on the choice of time coordinate, it holds for all
variations on the general line element.  Thus, the redshift is
infinite if the emitter is located at $r_1 = 2 M$, even if the
coordinate system allows for a crossing of the event horizon in finite
time.

\section{Vierbein Form of the General Line Element}
\label{vierbeins}

The general line element expresses all the metric degrees of freedom
that are possible by changing the time coordinate.  However, some
degrees of freedom lie ``beneath'' the metric; these are often
expressed as objects called \emph{vierbeins} $h^a_\mu$, which are like
a ``square-root'' of the metric.  The equivalence principle allows us
to assign an orthonormal frame of vectors (whose metric is just the
flat spacetime metric $\eta_{ab}$) to each point of spacetime, and the
vierbeins take us between the orthonormal frame and the full metric of
GR:
\beann
g_{\mu\nu} =h^a_\mu h^b_\nu \eta_{ab}\quad \to \quad \sqrt{-\det g} = \det h.
\eeann
Writing things out in terms of the vierbeins gives us not only the
general line element, but also lets us express the
\emph{isometries}---coordinate
transformations that do not change the form of the
metric.

We write in matrix form the vierbeins associated with the most general
spherically symmetric and stationary spacetime as
\beann
\left[ h^a_\mu \right] = \left( \begin{array}{cccc} g_1(r) & f_2(r) & 0 & 0 \\
g_2(r) & f_1(r) & 0 & 0 \\ 0 & 0 & r & 0 \\ 0 & 0 & 0 & r \sin \theta 
\end{array} \right)
\eeann 
so that 
\bea ds^2 = (g^2_1 - g^2_2) du^2 &+& 2 (f_1 g_2 - f_2
g_1)du dr \nonumber \\ &-& (f^2_1 - f^2_2) dr^2 - r^2 d\Omega^2 .
\eea 
(This is the way we first obtained the general line
element\cite{fk03a}.) \eq{GLE} follows from the identifications
\beann \chi = g^2_1 - g^2_2, \
\ B = f_1 g_2 - f_2 g_1, \ \ C = f^2_1 - f^2_2, 
\eeann 
and 
\bea f_1
g_1 - f_2 g_2 = 1.
\label{gauge2}
\eea
Einstein's equations then give \eq{chidef}, and \eq{gauge2} leads directly
to \eq{gauge}, so at first glance
it appears that these two conditions on the four
free functions leave two free functions to be specified. 
Nevertheless, the condition $C \chi + B^2 = 1$ arises from
$f_1 g_1 - f_2 g_2 = 1$, so that the metric still only contains one
free function.  The two free vierbein functions are both required to
determine the metric, so individually they represent isometric degrees
of freedom---a given expression for $B$ corresponds to a great many
choices of vierbein fields.

\end{appendix}

\bibliography{AmJP}

\end{document}